\documentclass[10pt]{article}

\usepackage{subcaption}

\usepackage[utf8]{inputenc}
\usepackage[T1]{fontenc}
\usepackage{lmodern}

\usepackage{amsmath,amssymb,dsfont}
\numberwithin{equation}{section}
\usepackage{microtype}
\usepackage{graphicx,tikz,pgfplots}
\graphicspath{{images/}}
\pgfplotsset{compat=newest}
\usepackage[hyperref,amsmath,thmmarks]{ntheorem}
\usepackage{aliascnt}
\usepackage[a4paper,centering,bindingoffset=0cm,marginpar=2cm,margin=2.5cm]{geometry}
\usepackage[pagestyles]{titlesec}
\usepackage[font=footnotesize,format=plain,labelfont=sc,textfont=sl,width=0.75\textwidth,labelsep=period]{caption}

\usepackage[backend=bibtex,maxnames=15]{biblatex}
\DefineBibliographyStrings{english}{%
	backrefpage = {cited on page},
	backrefpages = {cited on pages},
}

\title{Computed Origami Tomography}
\author{Axel Kittenberger$^1$, \, Leonidas Mindrinos$^1$ \, \mbox{and} \,  Otmar Scherzer$^{1,2}$ }
\date{}

\DeclareFieldFormat[report]{title}{``#1''}
\DeclareFieldFormat[book]{title}{``#1''}

\titleformat{\section}[block]{\large\sc\filcenter}{\thesection.}{0.5ex}{}[]
\titleformat{\subsection}[runin]{\bf}{\thesubsection.}{0.5ex}{}[.]

\usepackage[pdftex,colorlinks=true,linkcolor=blue,citecolor=green,urlcolor=blue,bookmarks=false]{hyperref}

\newaliascnt{proposition}{lemma}

\aliascntresetthe{proposition}

\newaliascnt{corollary}{lemma}

\aliascntresetthe{corollary}

\newaliascnt{example}{lemma}

\aliascntresetthe{example}

\newaliascnt{theorem}{lemma}

\aliascntresetthe{theorem}

\theorembodyfont{\normalfont}
\newaliascnt{definition}{lemma}
\newtheorem{definition}[definition]{Definition}
\aliascntresetthe{definition}

\newaliascnt{assumption}{lemma}

\aliascntresetthe{assumption}

\theoremstyle{nonumberplain}
\theoremseparator{:}
\theoremheaderfont{\normalfont\itshape}

\newtheorem{remark}{Remark}

\theoremsymbol{\ensuremath{\square}}

\newcommand{\R}{\mathds{R}}

\let\RE\Re
\let\Re=\undefined
\DeclareMathOperator{\Re}{\RE e}
\let\IM\Im
\let\Im=\undefined
\DeclareMathOperator{\Im}{\IM m}

\newcommand{\norm}[1]{\left\|#1\right\|}
\newcommand{\set}[1]{\left\{#1\right\}}

\let\ii\i
\renewcommand{\i}{\mathrm i}

\newcommand{\vx}{\vec x}
\newcommand{\vy}{\vec y}
\newcommand{\vz}{\vec z}
\newcommand{\vp}{\vec y}

\begin{document}

\maketitle
\thispagestyle{empty}
\begin{center}
\footnotesize E-mails: \href{mailto:axel.kittenberger@univie.ac.at}{axel.kittenberger@univie.ac.at},  \href{mailto:leonidas.mindrinos@univie.ac.at}{leonidas.mindrinos@univie.ac.at}, \href{mailto:otmar.scherzer@univie.ac.at}{otmar.scherzer@univie.ac.at}

\footnotesize $^1$Faculty of Mathematics, University of Vienna, Oskar-Morgenstern-Platz 1, 1090 Vienna, Austria

\footnotesize $^2$Johann Radon Institute for Computational and Applied Mathematics (RICAM), Altenbergerstraße 69, 4040 Linz, Austria
\end{center}

\begin{abstract} 
In this paper, we provide assembly instructions for an easy to build experimental setup in order to gain practical experience with tomography. In view of this, this paper can be seen as a complementary work to excellent mathematical textbooks at an undergraduate level concerned with the basic  mathematical principles of tomography. Since the setup uses light for tomographic imaging, the investigated objects need to be light transparent, like origami figures. 
In case the reader wants to experiment with computational tomographic reconstructions without 
assembling the device, we provide a database of several objects together with their tomographic measurements and a publicly available software. Moreover, recent advances in Cryo-imaging enabled three-dimensional high-resolution visualization of single particles, such as for instance viruses. To exemplify, and experience, single particle Cryo-electron microscopy we provide an advanced assembly, which we use to generate data simulating a Cryo-recording. We also discuss some of the major practical difficulties for reconstructing particles from Cryo-microscopic data. 
\end{abstract}
\section{Introduction}

Since the development of the first medically useable X-ray computerized tomography (CT) scanner, for which Allan M. Cormack and Godfrey N. Hounsfield received the Nobel Prize in Physiology or Medicine in 1979 \cite{Nobel79}, Radiology and non-destructive material testing has been revolutionized.

CT scanners allow ``seeing'' inside an object in a non-destructive way: In medical X-ray CT, the scanner is rotated around an axis and for every rotation angle an X-ray image is recorded. All recorded X-ray images are put together with a powerful computer, resulting in a three-dimensional (3D) representation of the interior of the patient.  The word tomography originates from the Greek word \emph{tomos}, which means slice or cross-section. This term might be rather misleading since nowadays the word tomography is used as a synonym for non-destructive 3D imaging based on any sort of illumination, such as light, magnetic currents or waves in general. While previously X-ray scanners reconstructed the body slice-by-slice (that is 2D by 2D), today a full 3D reconstruction can be performed, for instance with Spiral-CT or Cone-beam CT, without ``slicing'' the specimen. A curious application of these sophisticated algorithms is the scanning of tree logs
 \cite{GiuKatUrs16}.

The mathematical foundations for combining the X-ray images into a 3D volume, where laid down as early as 1917 by Johann Radon \cite{Rad17}. Nowadays, with increasingly sophisticated scanners, which allow for faster data acquisition and exposing the patients to less radioactive dose, also the mathematical sophistication of the algorithms increased (see for instance \cite{Kat02}). 

\begin{figure}[h]
\centering
  \begin{subfigure}{.35\textwidth}
        \centering
        \includegraphics[width=.8\linewidth]{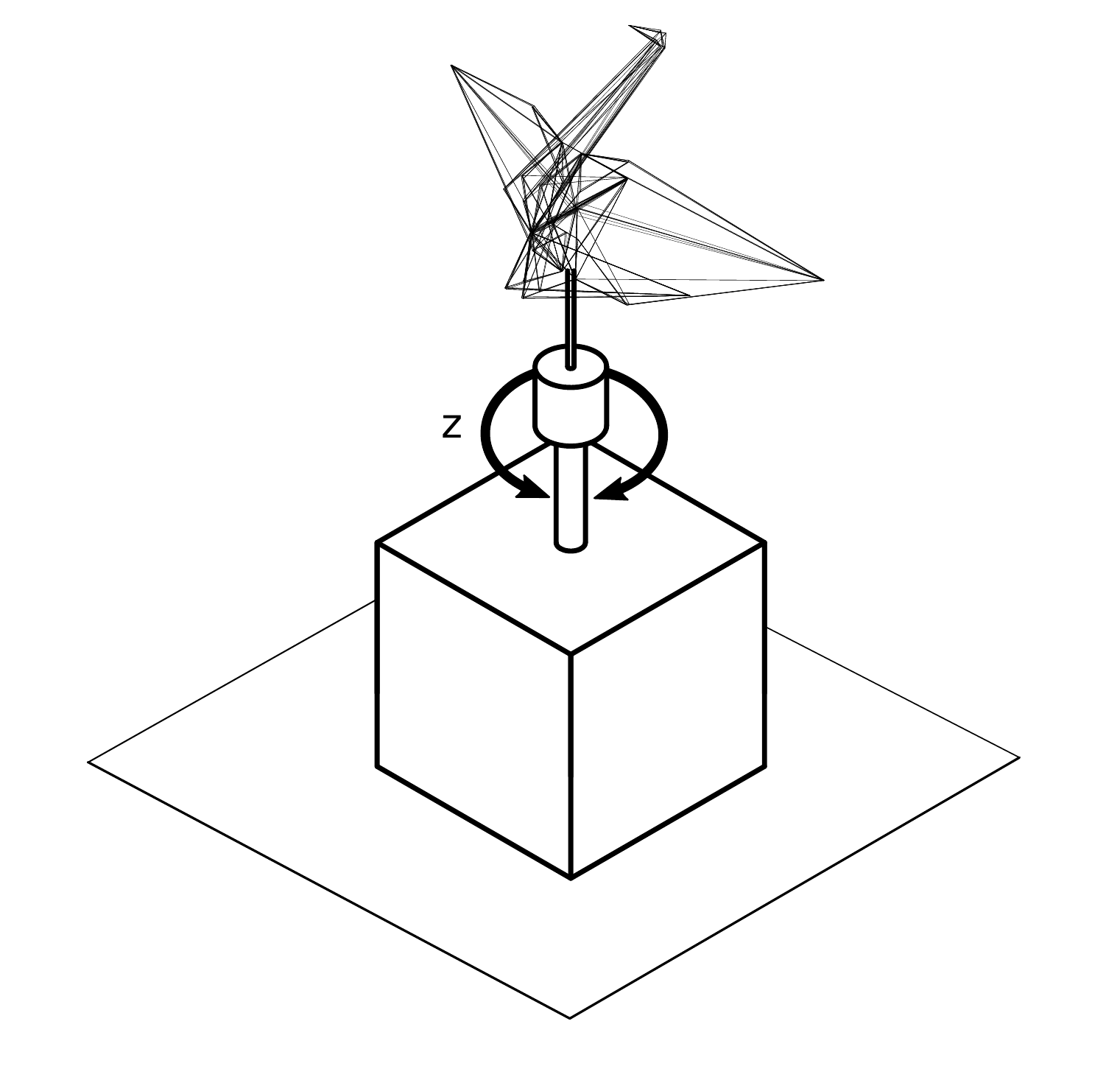}
    \end{subfigure}%
    ~ 
    \begin{subfigure}{.55\textwidth}
        \centering
        \includegraphics[width=.9\linewidth]{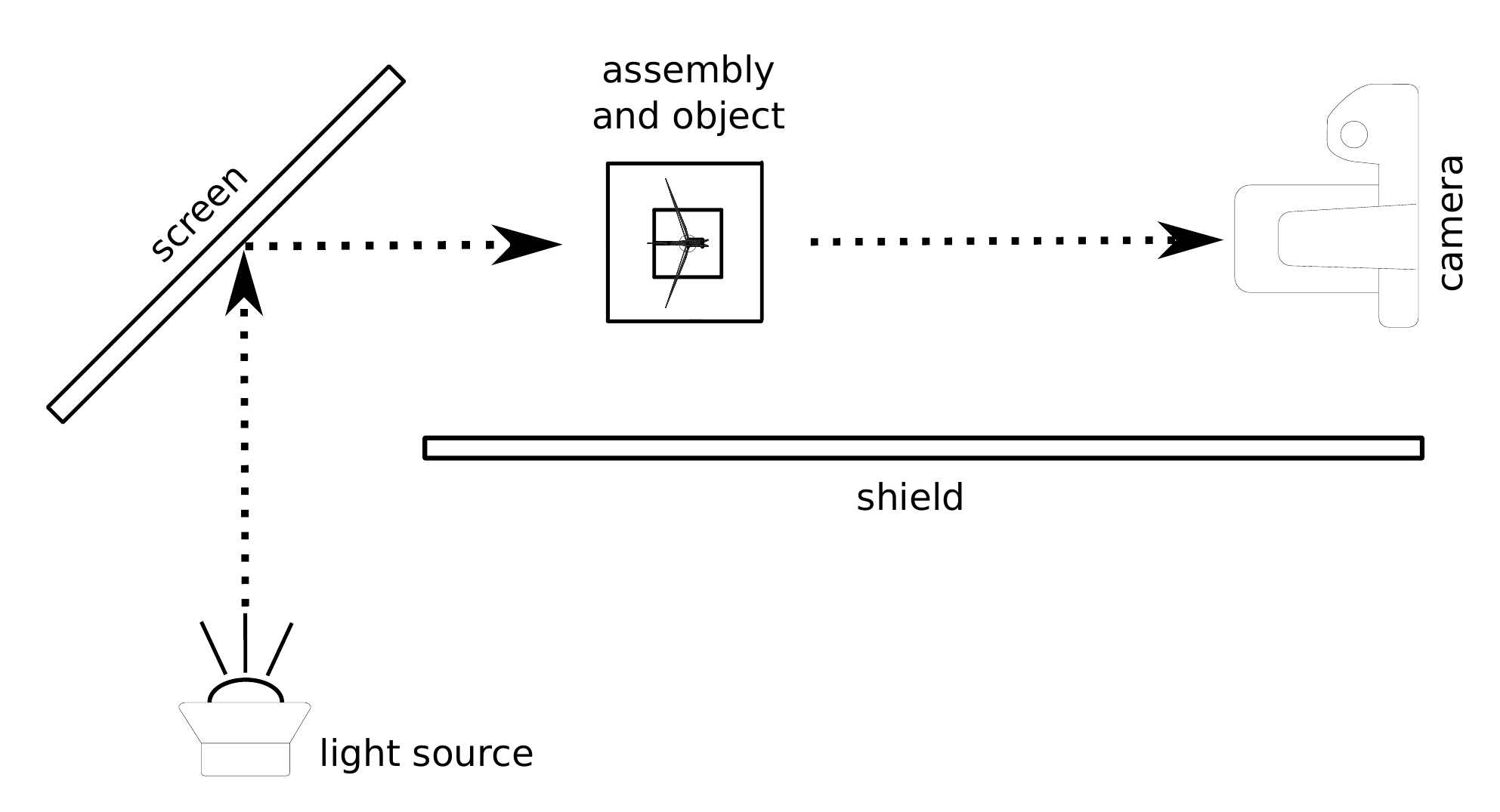}
    \end{subfigure}
    \caption{A stepper motor rotates the origami (left) and the proposed imaging system (right).}
\label{fig:opto}
\end{figure}

It belongs now to history that Allan M. Cormack built in 1963 a prototype CT scanner which costed approximately 100 USD \cite{Vau08}.  A fabulous investment! In this paper we give instructions for  crafting a scanner, which is based on \emph{optical} illumination and detection, respectively, and allows to exemplify a standard X-ray scanner. This system avoids dangerous X-ray radiation and thus can be safely used in training courses at high-schools or universities. The basics of this system are shown in \autoref{fig:opto} and are also explained in an educational video (in German) \cite{Sch_youtube}. 

Since we are using optical illumination the object under consideration needs to be transparent in the visible spectrum, equivalently to the human body, which is transparent to X-rays in CT.  Thus,  we propose to use origami figures out of cellulose acetate (CA) thin sheets, which are indeed optically transparent. Note that the attenuation coefficient of CA is around $1$mm$^{-1}$ in the visible range \cite{OreHokLepKam20}. For example, a breast tissue with diameter around $25$cm has attenuation coefficient approximately $0.4$cm$^{-1}$ (using CT) \cite{CheLonRigZanPel10}. Thus, a $15-20$mm thick origami object has  a similar (attenuation coefficient) $\times$ (thickness) factor, which tells us that our origami experiment is a realistic imaging scenario for CT imaging.
 
A fascinating new direction for computerized tomography is super-resolution  Cryo-imaging (the term ``Cryo'' refers to immobilizing a sample specimen by freezing it at very low temperatures). We all have been affected by the spread of the 2019-nCoV virus. At the beginning of 2020, we where already shown the first 3D high-resolution images of the Corona virus  \cite{ParWalWanSauLi19, WraWanCorGolHsi20}, which where produced with Cryo-electron microscopy (Cryo-EM). Jacques Dubochet, Joachim Frank and Richard Henderson were awarded the Nobel Prize in Chemistry in 2017 for the revolutionary work on immobilizing samples at low temperatures  \cite{Nobel17}. Aside from the chemical fascination concerned with freezing the specimen, there is an intriguing mathematical aspect of microscopy, which is for instance outlined in the special issue on Cryo-EM in the journal Inverse Problems \cite{SinSch20}.

\begin{figure}
\centering
   \begin{subfigure}{.45\textwidth}
        \centering
        \includegraphics[width=.6\linewidth]{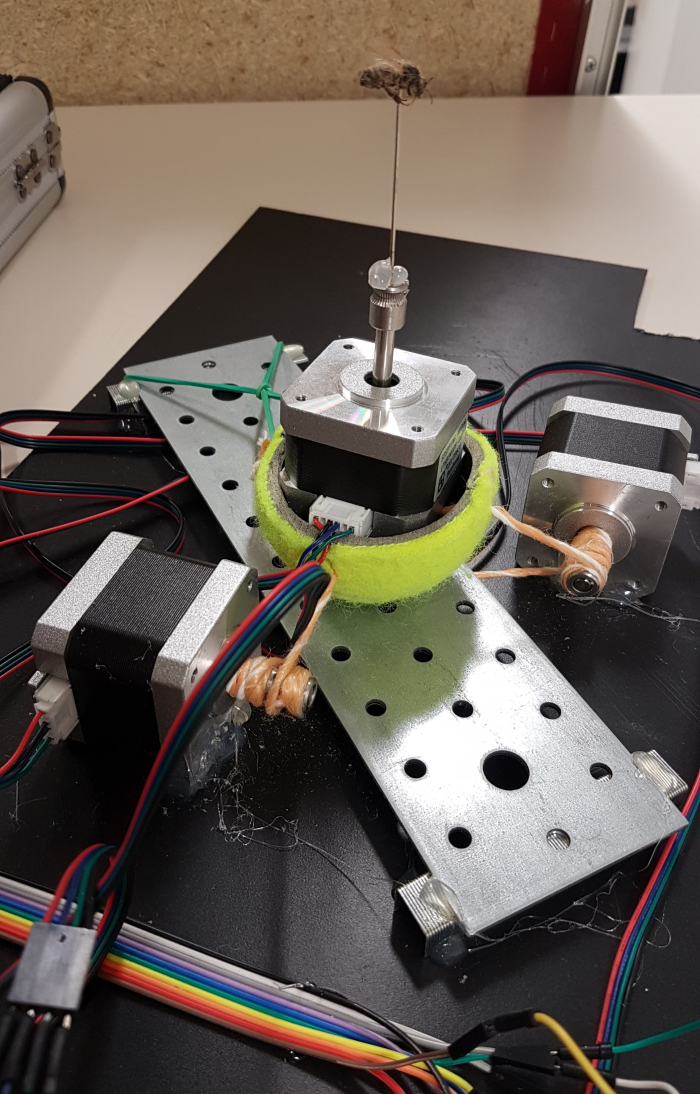}
    \end{subfigure}
    ~
  \begin{subfigure}{.45\textwidth}
        \centering
        \includegraphics[width=1\linewidth]{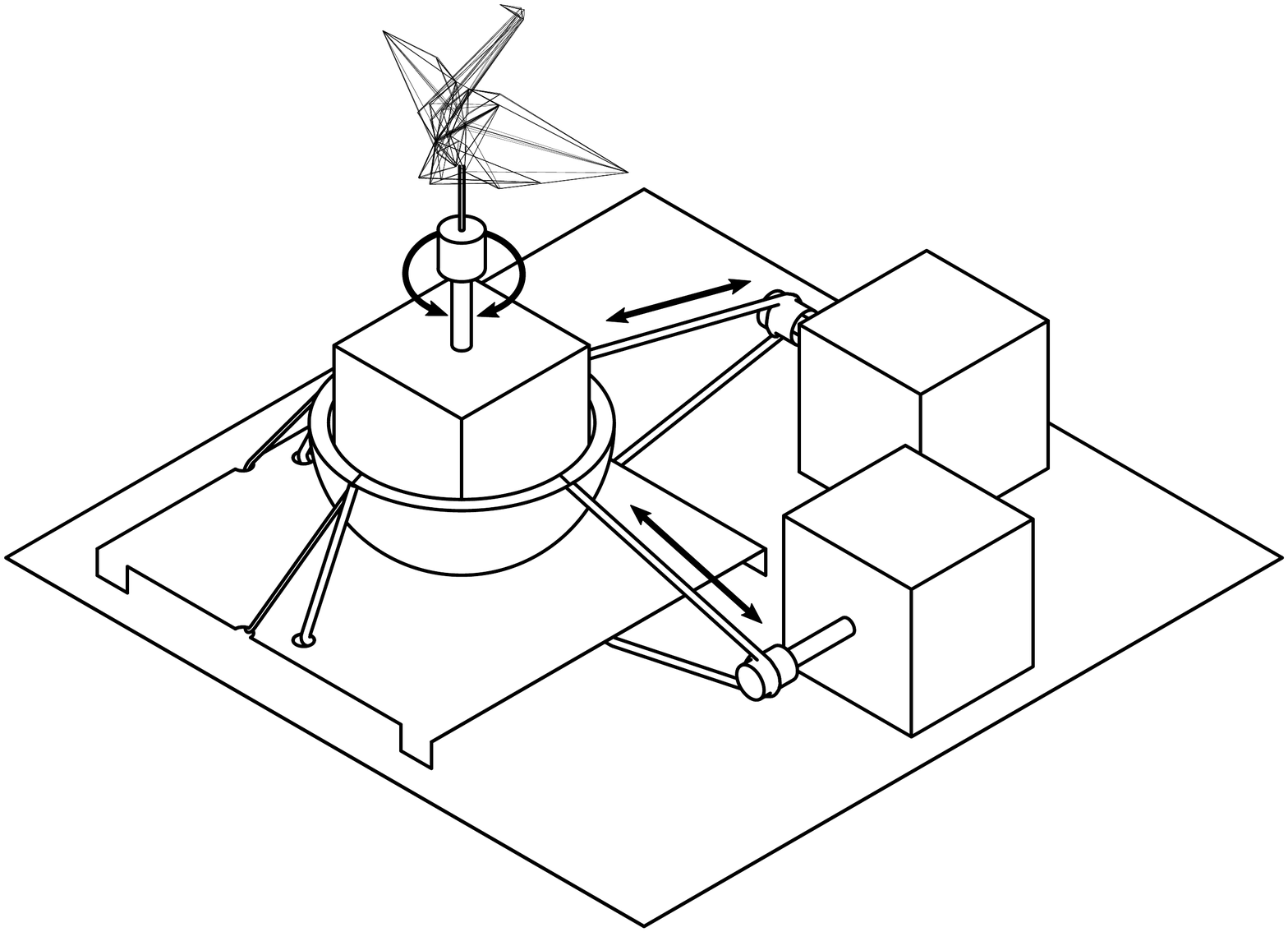}
    \end{subfigure}%
    \caption{The assembly (left) that was build to imitate the different orientations of the object and its isometric view (right).}
\label{fig:wobbling}
\end{figure}

In Cryo-imaging, we consider a sample of identical particles. These particles can be for instance viruses or molecular clusters in a cell or a cell-complex. Opposed to rotating the  biological sample and recording X-ray images, the whole sample is exposed to only one X-ray illumination. 
Cryo-microscopic imaging is based on the assumption that the recorded 2D image, called a \emph{micrograph}, contains numerous X-ray images of particles at different rotation angles.  Thus, in addition to tomographic imaging, Cryo-imaging also requires to localize the particles and the corresponding angular rotations, relative to the imaging direction. In order to exemplify such samples, we use an advanced assembly of the experimental tomographic setup, which, in addition, allows a ``wobbling'' rotation (see \autoref{fig:wobbling}). We use the recorded data to simulate a micrograph (called sample image), see \autoref{fig:micrograph} and \autoref{fig:circles}, which looks remarkably similar to real Cryo-EM data.

This paper builds on training courses, which we have offered to students at the age of 17-18, and it consists of the following parts:
\begin{enumerate}
\item We explain the physics of X-ray tomography, which is the Beer's law. This law is based on the assumption that X-rays travel along lines through the body. This is not completely true and modern CT scanners compensate for ``beam-widening''.  Our presented experiments use light illumination and we also assume that light travels along straight lines through the light transparent object. The theory of light-ray propagation is completely analogous to the theory of X-rays, and in Optics it is referred as the \emph{geometric optics} approximation (see \cite{Her00}).

\item We present instructions for building a crafting device which can record digital pictures (images) of small 3D objects at different rotation directions. In addition, we provide an open source software for reconstructing the 3D object from the collected images.

\item Finally, we give a short exposition on Cryo-imaging and we give again building instructions for a rotation device that can be used for illustrating Cryo-EM data.

\end{enumerate}
Typically in courses for young students we also present the Mathematics of the reconstruction formulas. However, here for the sake of a focused presentation we leave it out and we refer for instance to \cite{Fee10}.

\section{Modeling X-Ray Tomography}\label{sec:ray}

Standard X-ray tomography is based on the following two assumptions that 
\begin{enumerate}
\item an X-ray propagates straight (along a line) through the body (this is why it is named ray) and 
\item during propagation it loses intensity (that is the number of traveling particles), because particles are absorbed by the medium. 
\end{enumerate}
Of course, these two assumptions are not completely true in practice, and modern CT-scanners compensate for many effects which appear during imaging. The quantity of how many particles are absorbed per unit length is visualized in X-ray CT and is used for diagnostic purposes.

\subsection{Computerized transverse axial (CTA) scanning}\label{sec:cta}
\begin{figure}[bht]
        \centering
        \includegraphics[width=.5\linewidth]{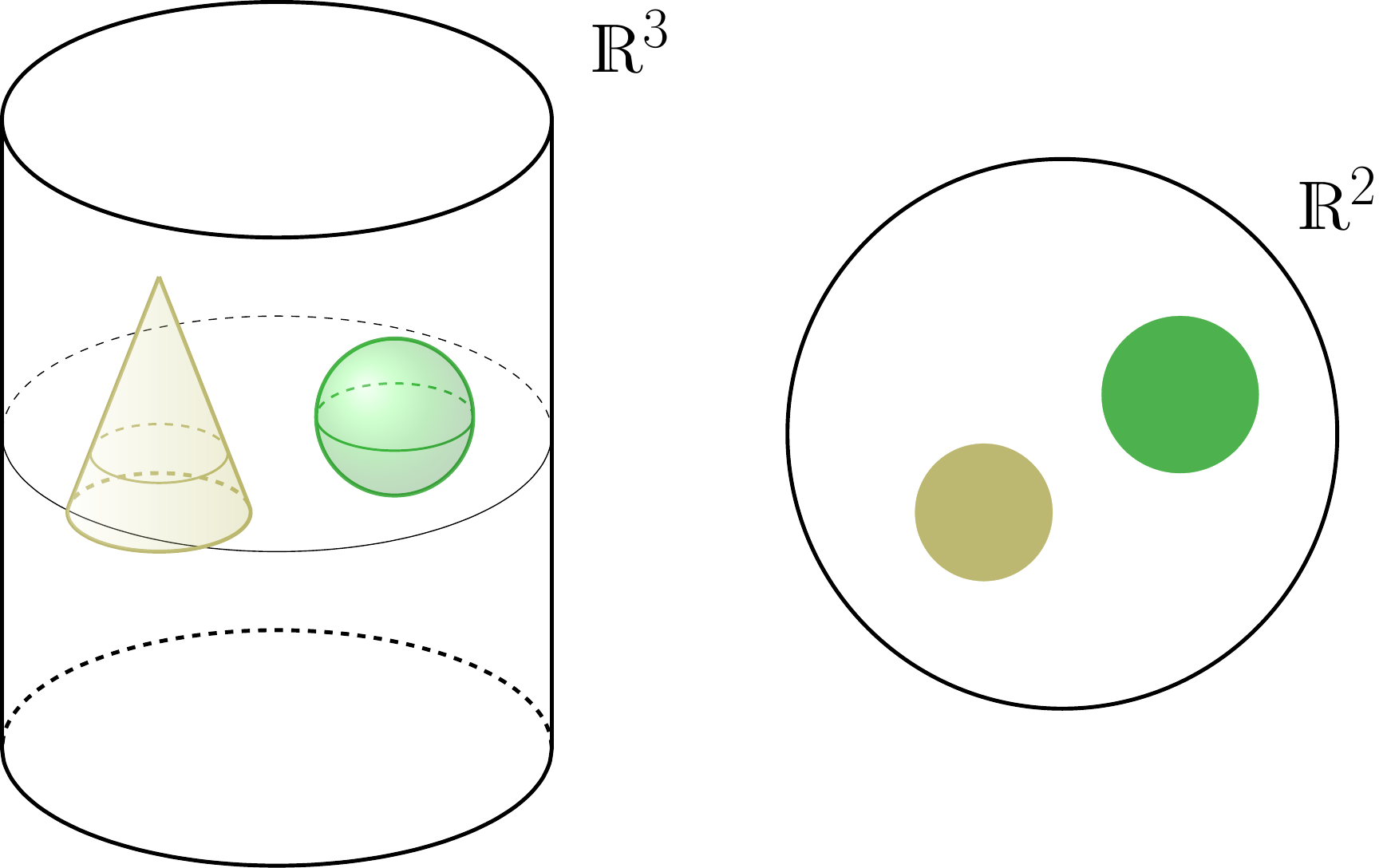}
        \caption{\label{fig:slice} The CTA technique is based on dividing the 3D object in a series of slices.}
\end{figure}

In what follows we consider X-ray CT, as it was first implemented with \emph{computerized transverse axial (CTA)}  scanners.  
There, the 3D object is visualized by scanning slice by slice (thus performing 2D tomography for every slice) orthogonal to the rotation axis, and then by combining the slices we obtain the 3D volume (see \autoref{fig:slice}).

Let us start by modeling the absorption of an X-ray  propagating along a straight line in a slice from a point $\vx =(x_1,x_2) \in \R^2$ to $\vy =(y_1,y_2) \in \R^2$. The direction of the line is given by 
\begin{equation*}
 \vec{d} = \frac{\vy-\vx}{\norm{\vy-\vx}}, \text{ where } 
 \norm{\vy-\vx} = \sqrt{(y_1-x_1)^2+(y_2 -x_2)^2} 
\end{equation*}
denotes the distance between $\vx$ and $\vy$. If $\norm{\vy-\vx}=0$, then we set $\vec{d}=\vec{0}$, the 2D zero-vector.

For every point $\vz =(z_1,z_2)$ on the line between $\vx$ and $\vy$ we denote by $I(\vz)$ the intensity of the X-ray  at position $\vz$. Since $\vz$ lies between $\vx$ and $\vy$ and the X-ray  is traveling from $\vx$ to $\vy$ we can express 
\begin{equation*}
\vz = \vx + s \vec{d}, \text{ with } s \in (0,\norm{\vy-\vx}) 
\text{ denoting the distance between } \vz \text{ and } \vx.
\end{equation*}
Let us first assume that the medium is homogeneous, meaning that it is has the same  X-ray absorption properties everywhere. Then for any two points between $\vx$ and $\vy$,
\begin{equation} \label{eq:vz}
\vz_1 = \vx + s_1\vec{d} \text{ and } \vz_2 = \vx + s_2\vec{d}, \text{ with } 0 < s_1 < s_2 < \norm{\vy-\vx},
\end{equation}
the factor 
\begin{equation} \label{eq:absorpt}
  \frac{I(\vz_1)-I(\vz_2)}{I(\vz_1)} \frac{1}{s_2-s_1} = \mu,
\end{equation}
denotes the \emph{absorption coefficient} of the homogeneous material per unit length. 
The first term in \autoref{eq:absorpt} denotes the percentage of absorbed particles between $\vz_1$ and $\vz_2$, and therefore by dividing by $s_2-s_1$ we get the absorption per unit length. 
Note that $I(\vz_1)-I(\vz_2)$ is positive since $s_2 > s_1$ and the X-ray is propagating from $\vx$ to $\vy$ via $\vz_1$ and $\vz_2$. The term $\frac{I(\vz_1)-I(\vz_2)}{I(\vz_1)}$ is dimensionless, that is, it is a percentage, or in other words the relative loss of intensity. This means that
the absorption coefficient $\mu$ has units of inverse length. For X-rays, it is measured in inverse centimetre (cm${}^{-1}$).  

\begin{remark}
\autoref{eq:absorpt} makes sense only if the intensity $I(\vz_1)$ does not vanish (for a zero intensity the relative loss of intensity cannot be reasonably defined). As a consequence  $\mu \in [0,\infty)$.
\end{remark}

If the absorption is in-homogeneous, like in the human body, meaning that $\mu = \mu(\vz)$ is spatially varying, then for any two points $\vz_1$ and $\vz_2$ on a line between $\vx$ and $\vy$,  with parametrizations $s_1 < s_2$, respectively, analogously to  \autoref{eq:absorpt}, we find that the relative loss of intensity of the X-ray is given by
\begin{equation} \label{eq:aa}
  - \frac{I \left(\vx+s_2 \vec{d} \right)-I \left(\vx+s_1 \vec{d} \right)}{s_2-s_1} \frac{1}{I(\vz_1)} = 
  \frac{I(\vz_1)-I(\vz_2)}{I(\vz_1)} \frac{1}{s_2-s_1}. 
\end{equation}
If $\vz_2$ is close to $\vz_1$, then also $s_2$ is close to $s_1$ and the first term is nothing else than an approximation of the derivative of the function $s \in (0,\norm{\vy-\vx}) \to I \left(\vx+s \vec{d} \right)$. Thus, by taking the limit $s_2 \to s_1$ we get 
\begin{equation} \label{eq:ab}- \frac{I' (\vz_1)}{I (\vz_1)} = 
  - \frac{I' \left(\vx+s_1 \vec{d} \right)}{I \left(\vx+s_1 \vec{d} \right)} =: \mu \left(\vx+s_1 \vec{d} \right) =\mu(\vz_1).
\end{equation}
In other words, \autoref{eq:ab} is the generalization of \autoref{eq:absorpt} for in-homogeneous media. Then, by using the chain rule $(f \circ g)'(s) = f'(g(s))g'(s)$ for $f = \mbox{ln}$, 
the natural logarithm, and  $g=I$, 
\begin{equation} \label{eq:beer}
\mu(\vz_1)= - \frac{I'(\vz_1)}{I(\vz_1)} = -  (\mbox{ln} \circ I)'(\vz_1).
\end{equation}
This is know as \emph{Beer's law}:
\begin{definition}[Beer's law \cite{Bee52}] The intensity of a X-ray  and the absorption of the material are related by \autoref{eq:beer}. 
\end{definition}
Therefore, for an X-ray propagating from $\vx$ to $\vy$, we get from Beer's law that
\begin{equation} \label{eq:beer_int}
\begin{aligned}
\int_{0}^{\norm{\vy-\vx}} \mu\left(\vx+s \vec{d} \right) ds & = - \int_{0}^{\norm{\vy-\vx}} \frac{I'\left(\vx+s \vec{d} \right)}{I\left(\vx+s \vec{d} \right)} ds \\
& = - \int_{0}^{\norm{\vy-\vx}} (\mbox{ln} \circ I)'\left(\vx+s \vec{d} \right)) ds\\ 
&=   \mbox{ln} (I(\vx))-\mbox{ln} (I(\vy)).
\end{aligned}
\end{equation}

For imaging, one sends a ray with known intensity into the body, so one knows $I(\vx)$  at a point $\vx$ before it enters the body. The intensity of the ray is measured at  a point $\vy$ after the ray has passed through the body. Thus, the right-hand side of \autoref{eq:beer_int} is known for a pair of points $\vx$ and $\vy$ ``before'' entering and ``after'' exiting the body.

Typically, one makes the additional assumption that no particles of the ray are absorbed outside of a ball with radius $R$, or in other words that $\mu \equiv 0$ outside the  ball. In medical terms, $R$ describes the maximum thickness that a sample can have in order to be examined with the scanner. 

\begin{figure}[bht]
        \centering
        \includegraphics[width=.5\linewidth]{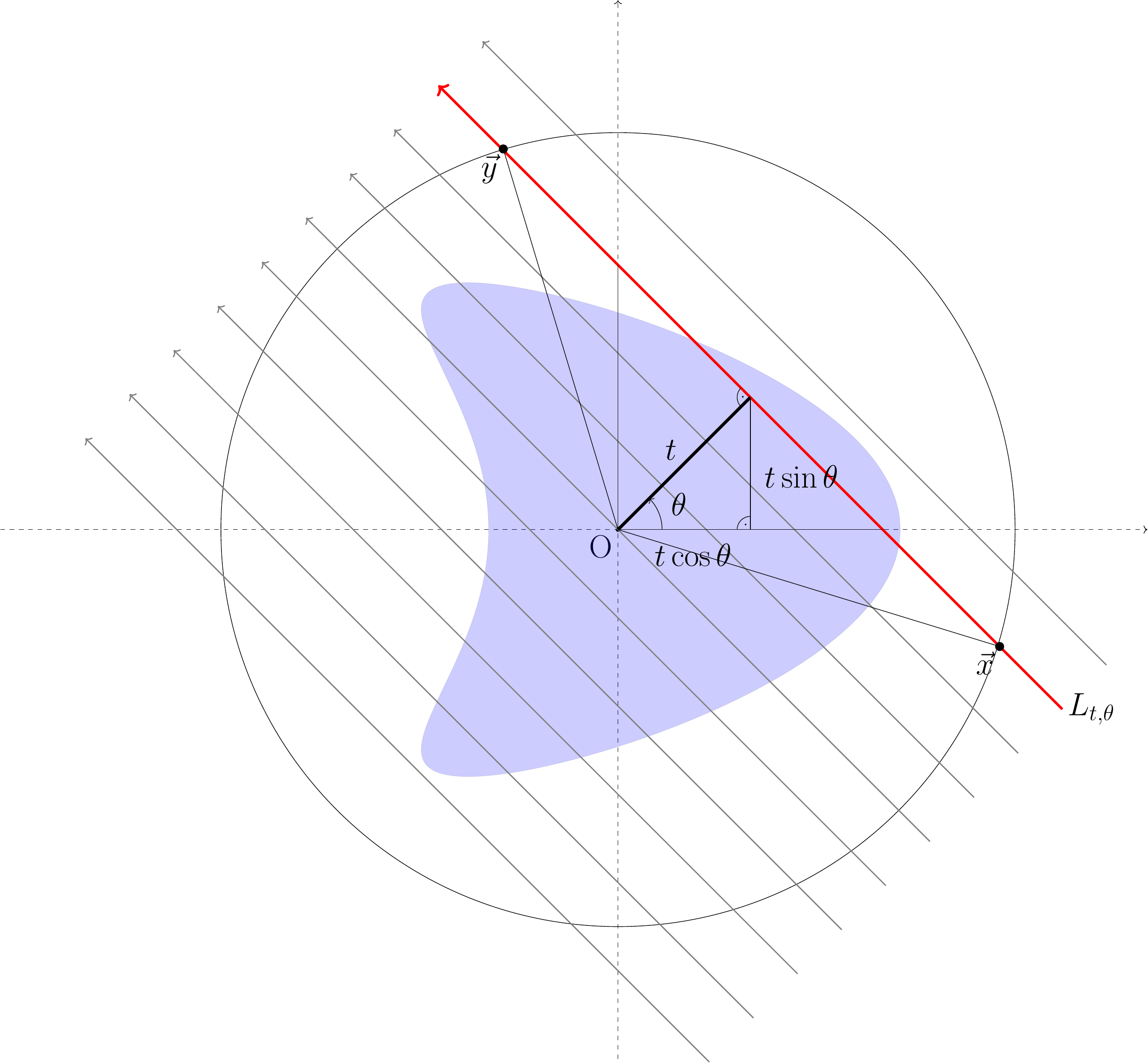}
        \caption{\label{fig:para} The parametrization of the line $L_{t,\theta}$.}
\end{figure}

We denote by $L_{t,\theta}$ the line, with a signed \emph{normal distance} $t \in \R$ from the origin and with orientation $\vec{v} = (-\sin \theta, \cos \theta)$, where $\pi-\theta \in [0,\pi)$ denotes the inclination of the line with respect to the horizontal axis (see \autoref{fig:para}). Note that the line is orthogonal to $\vec{v}^\bot = (\cos \theta, \sin \theta)$.
This means that $L_{t,\theta} = \set{ t \vec{v}^\bot + s \vec{v}: s \in\R }$.
If the line $L_{t,\theta}$ passes through $\vx$ and $\vy$, which are points outside of the body, then we can write the left-hand side of \autoref{eq:beer_int} as 
\begin{equation}\label{beer_final}
\mathcal{R} [\mu] (t,\theta) := \int_{L_{t,\theta}} \mu(\vz) d s (\vz) = 
\int_{-\infty}^{\infty} \mu\left(t \vec{v}^\bot +s \vec{v} \right) ds.
\end{equation}
The function $\mathcal{R} [\mu]$ is called the \emph{Radon-transform}, sometimes also called the \emph{X-ray transform}  in $\R^2,$ of $\mu$.

\subsection{Mathematics of CTA scanners}
The mathematical problem of X-ray computerized tomography consists in calculating the spatially varying absorption coefficient $\mu$ from the measured intensities of X-ray beams, which have propagated through the body. 
This problem is mathematically equivalent to estimating a compactly supported function $\mu$ from the knowledge of $\mathcal{R}[\mu]$.  Note that the Radon-transform is a function of a signed distance and an angle, while $\mu$ is a function of two spatial variables. 
The Radon-image, that is the set $
 \set{\mathcal{R} [\mu] (t,\theta) : (t,\theta) \in \R \times [0,\pi)},$
is often called \emph{sinogram}.

Elementary expositions of mathematical aspects, such as calculating $\mathcal{R}[\mu]$ analytically for simple functions $\mu$ and also the inverse direction of calculating $\mu$ from $\mathcal{R}[\mu]$, can be found for instance in \cite{Fee10}. Aside from this elementary textbook much deeper mathematical analysis can be found in \cite{Kuc13,Nat01,NatWue01,Sch15}.

\section{The Origami Scanner}\label{sec_origami_scanner}
In this section we describe our origami scanner (see \autoref{fig:opto}) consisting of three
essential parts, (1) an illuminated screen, (2) a stepper motor and (3) a digital camera.
Transparent origamis folded from a thin cellulose acetate sheet of size $50 \times 50$mm, are 
ideal test-objects for our setup.
Folding instructions of the origami crane used in this paper are given in \cite{Crane}. We scanned also other, even non-transparent, objects, and we achieved remarkably good reconstructions too. Several samples can be found in our database (available at \url{https://csc1.gitlab.io/otomo-samples/}). 

The test object is either pinned on a needle or fastened with hot-melt adhesive onto a stick which is in turn glued with hot-melt adhesive to a pulley. The later, in turn, is secured with a screw onto the axis of a standard NEMA 17 stepper motor, which we call the $z$-\emph{stepper}.
It is either clamped in a vise or glued on a plate guaranteeing a stable vertical rotation axis; See the left picture of \autoref{fig:opto} for an illustration of the assembly.

The whole setup with the light path is shown in the right picture of \autoref{fig:opto}. There exist two possibilities for the illuminated screen. A laptop screen illuminates directly the 3D object or a light bulb (which should be the only light source in a completely darkened room) illuminates indirectly the object through a white screen. The light from the screen passes through the object and reaches the camera. In case of using a light bulb an extra shield is placed between the light source and the object to minimize reflective interference from direct illumination.

We control the stepper via an A4988 stepper motor driver with an Arduino single-board controller (further called micro-controller). A PC is used to control both the micro-controller and the camera. The PC and the micro-controller communicate with a simple, serial, single character protocol over a USB cable. It resets the stepper into its zero-position if the connection with the PC turns idle.  To operate the camera through the PC, via a USB cable, we use the ``Canon Hack Development Kit (CHDK)'' \cite{CHDK}.  The source code of all created scripts is available at \url{https://csc1.gitlab.io/otomo-tk/}.

Since a recording encompasses hundreds of digital images, the ``remote-shoot'' function of CHDK is used to load the pictures directly to the PC bypassing the SD card. For the camera, the use of a power supply instead of a battery is advised. 
In our case, the camera needs about 1.5 seconds for one picture and  we have to wait half a second between the shoots in order to avoid camera stalling due to overload. It takes approximately 15 minutes to record the $N=400$ images. The filenames of the saved data correspond to the imaging directions.
We observed that if we fully rotate the object $360^\circ$, and thus record double as many data as are actually required in a reconstruction process,  we get better results. We guess that this has to do with uncertainties in our  imaging system, which are statistically averaged by this oversampling.

\begin{figure}[t]
\centering
  \begin{subfigure}{.3\textwidth}
        \centering
        \includegraphics[width=.8\linewidth]{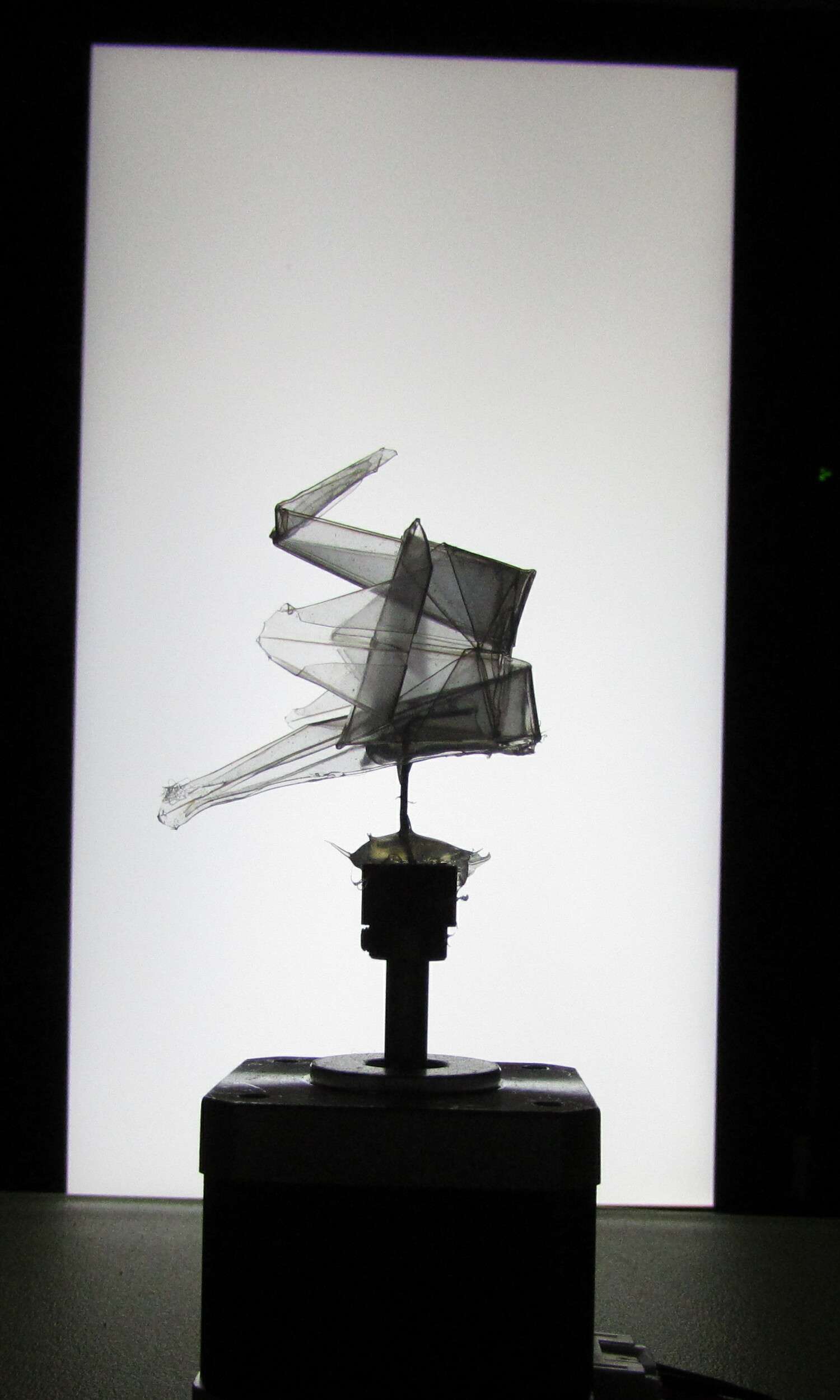}
     \end{subfigure}%
    ~ 
    \begin{subfigure}{.3\textwidth}
        \centering
        \includegraphics[width=.9\linewidth]{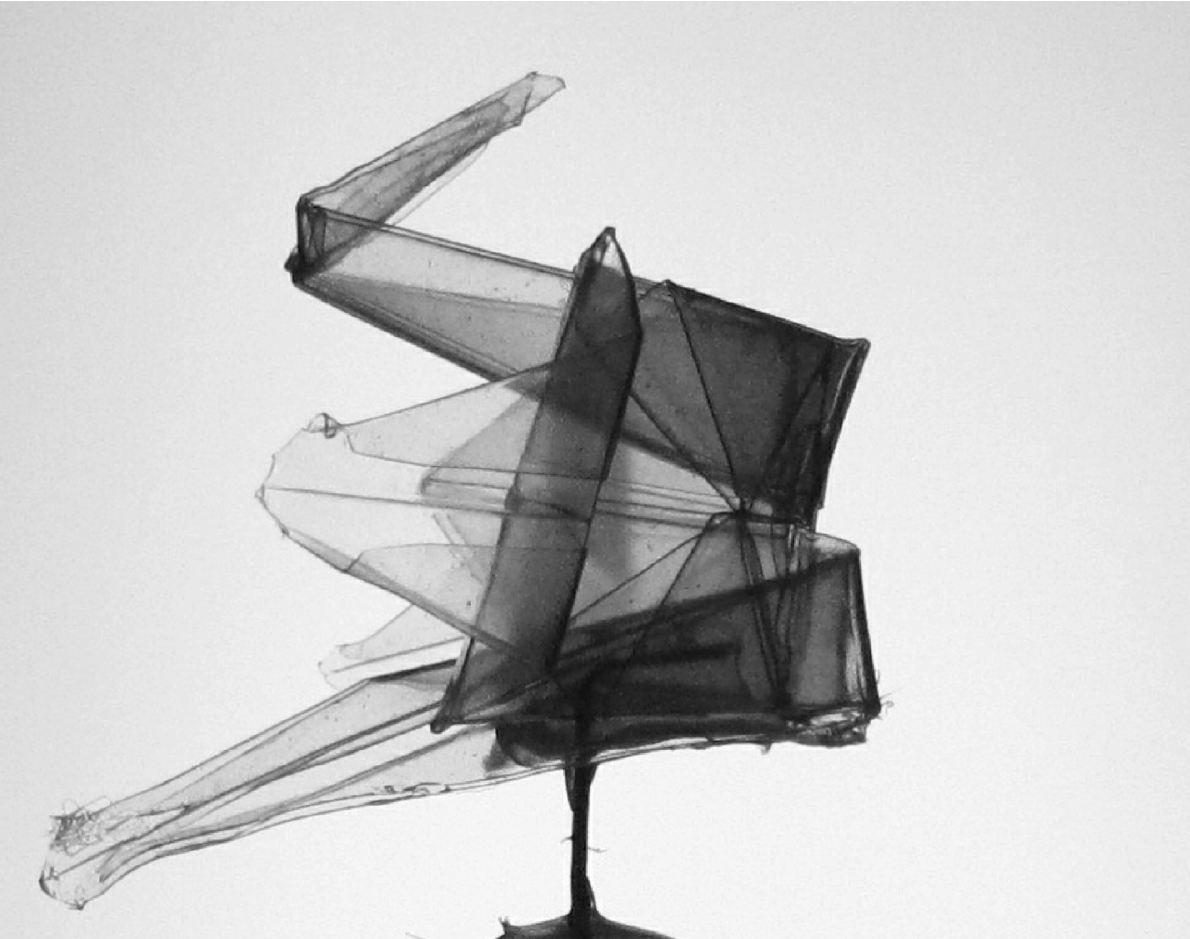}
    \end{subfigure}
      ~ 
    \begin{subfigure}{.3\textwidth}
        \centering
        \includegraphics[width=.9\linewidth]{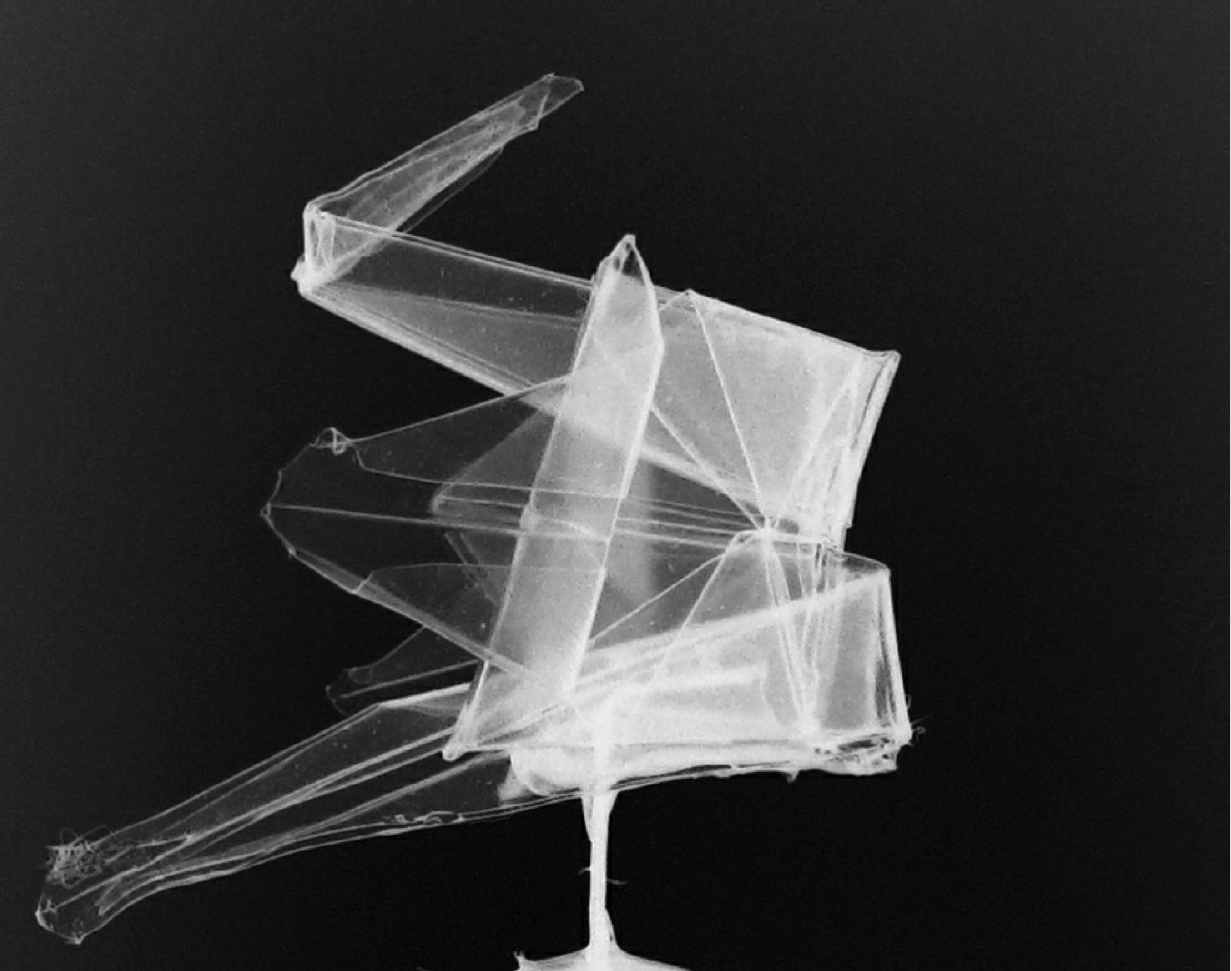}
    \end{subfigure}
    \caption{The original image (left), its cropped version (middle) and the converted final image (right).}
\label{fig:crane}
\end{figure}

\subsection{CT data simulation and reconstructions}
Before starting with the tomographic reconstruction, we have to post-process the recorded images. We start by aligning the rotation axis (that is the needle where the origami is fixated) in the vertical direction. Orthonormal to the axis we determine a symmetric region of interest and we cut out the rest. To ease this task we developed a \texttt{C++} application called ARTI (Accelerated Radon Transformation Interface), available for download at  \url{https://csc1.gitlab.io/arti/}, which provides a GUI to handle sequences of images. In the middle picture of \autoref{fig:crane}  we see the cropped image.

During exposure, a digital camera counts the numbers of arriving photons at every sensor pixel. Here, we just refer to luminous exposure and not to the generated voltage since the photon intensity is proportional to the intensity of the emitted electrons. Thus, in turn, the  voltage strength is proportional to the number of detected photons, and it correlates to the intensity value of a pixel at the recorded digital images (see \autoref{eq:pixel} below).

The represented intensities on a digital picture have to be in accordance with  human perception, which is proportional to the logarithm of the number of detected photons. For a grayscale image, the typical pixel value is an 8-bit data value, with range from $0$ (black) to $255$ (white). The used camera has a CMOS sensor and following \cite{DoYoo16}  we see that the displayed intensity value $V(\vp)$ at the pixel $\vp$ of the digital photo is given by
\begin{equation}\label{eq:pixel}
V(\vp) = 255 \left(\frac{I(\vp)}{I_{max}} \right)^{1/\gamma},
\end{equation}
where $\gamma >1$ is a fixed parameter
and $I_{max}$ denotes the maximum possible light intensity.  The constant $\gamma$ (called the encoding gamma) is specified by the camera settings and its typical value is $2.2.$  The above equation is equivalent to \cite[Equation 3]{DoYoo16} since the photon intensity is directly proportional to the voltage and thus we can replace the voltage ratio with the intensity ratio.  
Taking the logarithm of \autoref{eq:pixel} and using its elementary properties we get
\begin{equation} \label{eq:gamma}
- \gamma \,\mbox{ln} \left(\frac{V(\vp)}{255} \right) = \mbox{ln} \left(\frac{I_{max}}{I(\vp)} \right).
\end{equation}

\begin{figure}
\centering
\includegraphics[width=.5\linewidth]{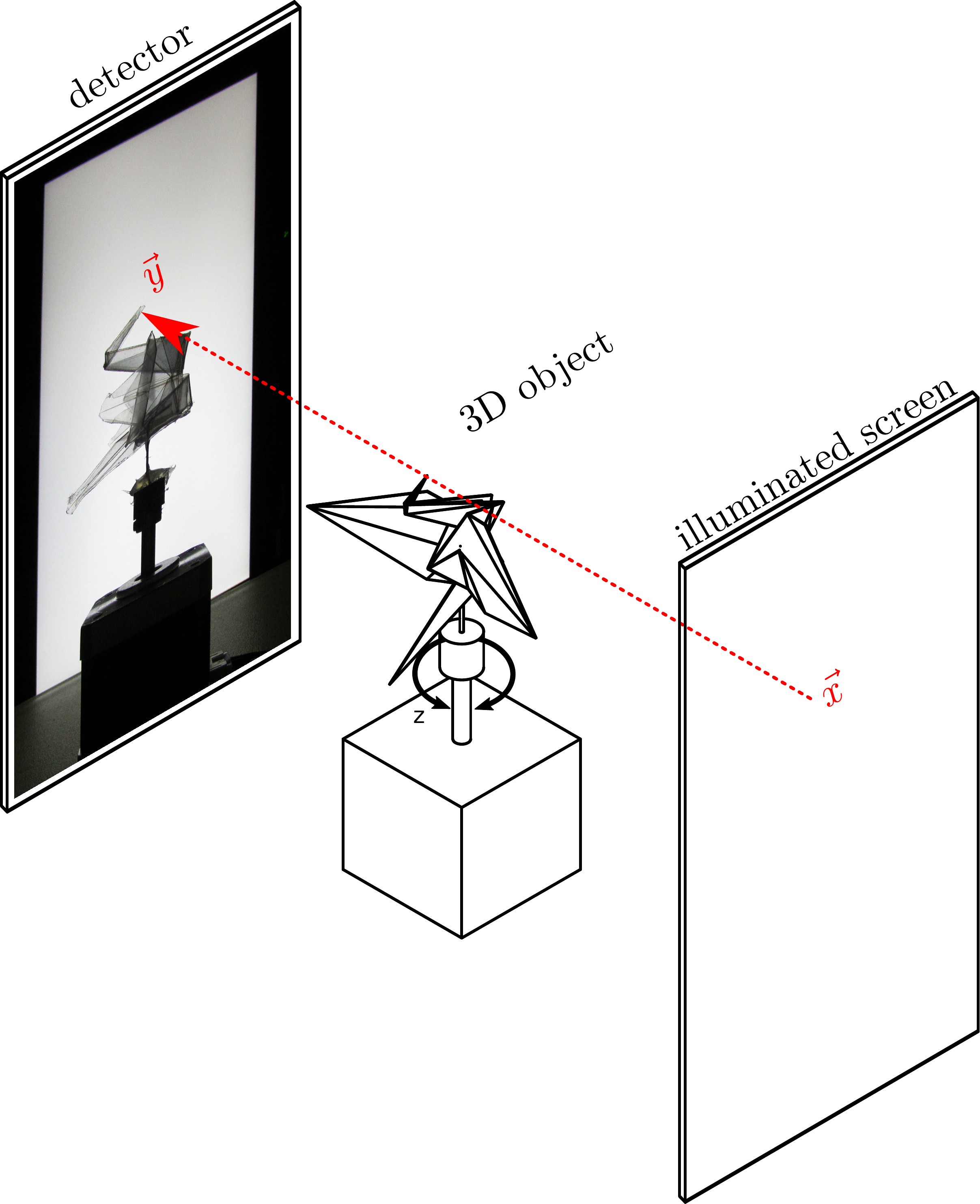}
\caption{The pixel value $V(\vy)$ at the image is related to the light intensity $I(\vy)$ (see \autoref{eq:pixel}) and consequently to the Radon data, meaning the integral along the line passing through $\vx$ and $\vy$ (see \autoref{eq:beer_int}).}
\label{fig:3dsetup}
\end{figure}

Let $\vx$ be a point on the white screen and let $\vy$ be a sensor point at the digital camera; See \autoref{fig:3dsetup} for the setup, where for illustration purposes we omit the optics of the camera. The white screen represents the light source and thus we have maximum intensity $I_{max}$ there. Between the origami and the sensor there is no absorption (air). Then, from \autoref{eq:beer_int}  and \autoref{eq:gamma}, it follows that 
\begin{equation}\label{radon_image}
\mathcal{R}[\mu](t,\theta) =\mbox{ln} \left(\frac{I_{max}}{I(\vp)} \right) = - \gamma \,\mbox{ln} \left(\frac{V(\vp)}{255} \right),
\end{equation}
where $L_{t,\theta}$ is the line passing through $\vx$ and $\vy$.
Therefore, for Radon inversion, the recorded digital image $V$ has to be pre-processed as shown in the right-hand side of \autoref{radon_image}, meaning normalize, take the logarithm and scale. 

\begin{remark}
The above formula holds only if $V(\vy)>0,$ for all sensor points $\vy.$ A zero pixel value can occur at in-transparent parts of the object. In our case, this can happen for example at the region of the needle or the pulley. 
\end{remark}

The right hand side of \autoref{radon_image} can be simplified by approximating the logarithm by a linear function. Then, we obtain
\begin{equation}\label{radon_image2}
\mathcal{R}[\mu](t,\theta) \approx \gamma \, \left(1-\frac{V(\vp)}{255} \right) = \frac{\gamma}{255} (255 - V(\vp)).
\end{equation}
The term in the parenthesis, which represents the compliment of the image, can be easily  implemented with the \texttt{Matlab} function \texttt{imcomplement}. The scaling factor $\tfrac{\gamma}{255}$ can be neglected (linearity of the Radon transform) but then the reconstructed function has to be rescaled in order to approximate $\mu.$ In the following, we do not consider the scaling factor. In the right picture of \autoref{fig:crane}, we see the converted image. 

\begin{remark} 
One positive practical aspect of the approximation, \autoref{radon_image2}, is that it does not require $V(\vp)$ to be positive (in contrast to \autoref{radon_image}) to be well-defined. Indeed, since $V(\vp)$ is rounded in order to be represented in 8 bits, it can be zero rather frequently. Working with the original model, \autoref{radon_image}, would require to constrain $V$ away from zero, meaning use another approximation. 
\end{remark}
 
\begin{figure}
\centering
\includegraphics[width=.8\linewidth]{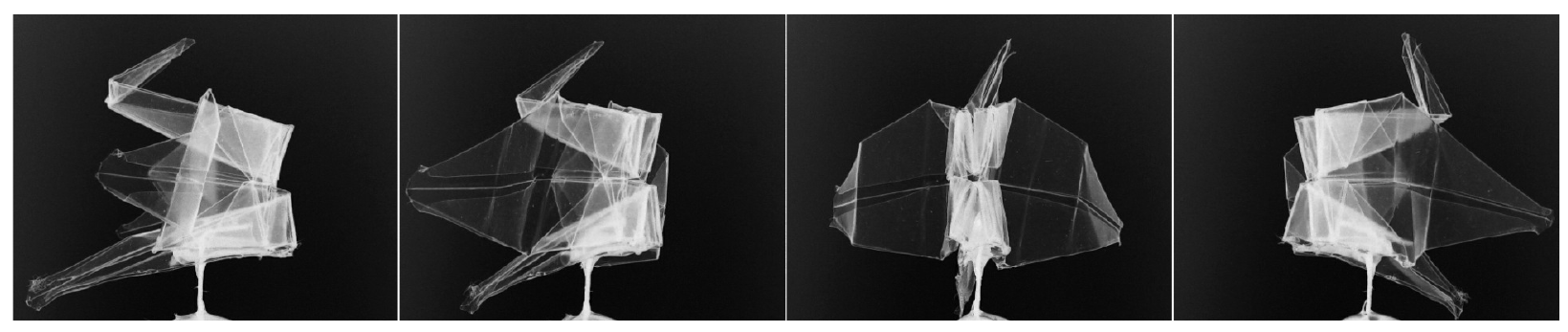}
\caption{Post-processed images (4 of total 400) of the 3D origami crane.}
\label{fig:postpro}
\end{figure}

The processed images have size of $765 \times 971\, (h \times w).$ Four of them are shown in \autoref{fig:postpro}. The resulted data have compact support in the image domain, and thus we are consistent with the definition of the Radon transform, \autoref{beer_final}, where we assumed that the absorption coefficient $\mu$ has compact support.
The simulated data with dimensions $765 \times 971 \times 400\, (h \times w \times N)$ are then the collection of the 2D object images.  The sinograms, equivalent to the 2D case, are the $765$ horizontal cross-sections of the volumetric data and thus have size $971 \times 400.$  This can be easily done in \texttt{Matlab} by rearranging the dimensions of the volumetric data.
\begin{figure}[bht]
\centering
\includegraphics[width=.8\linewidth]{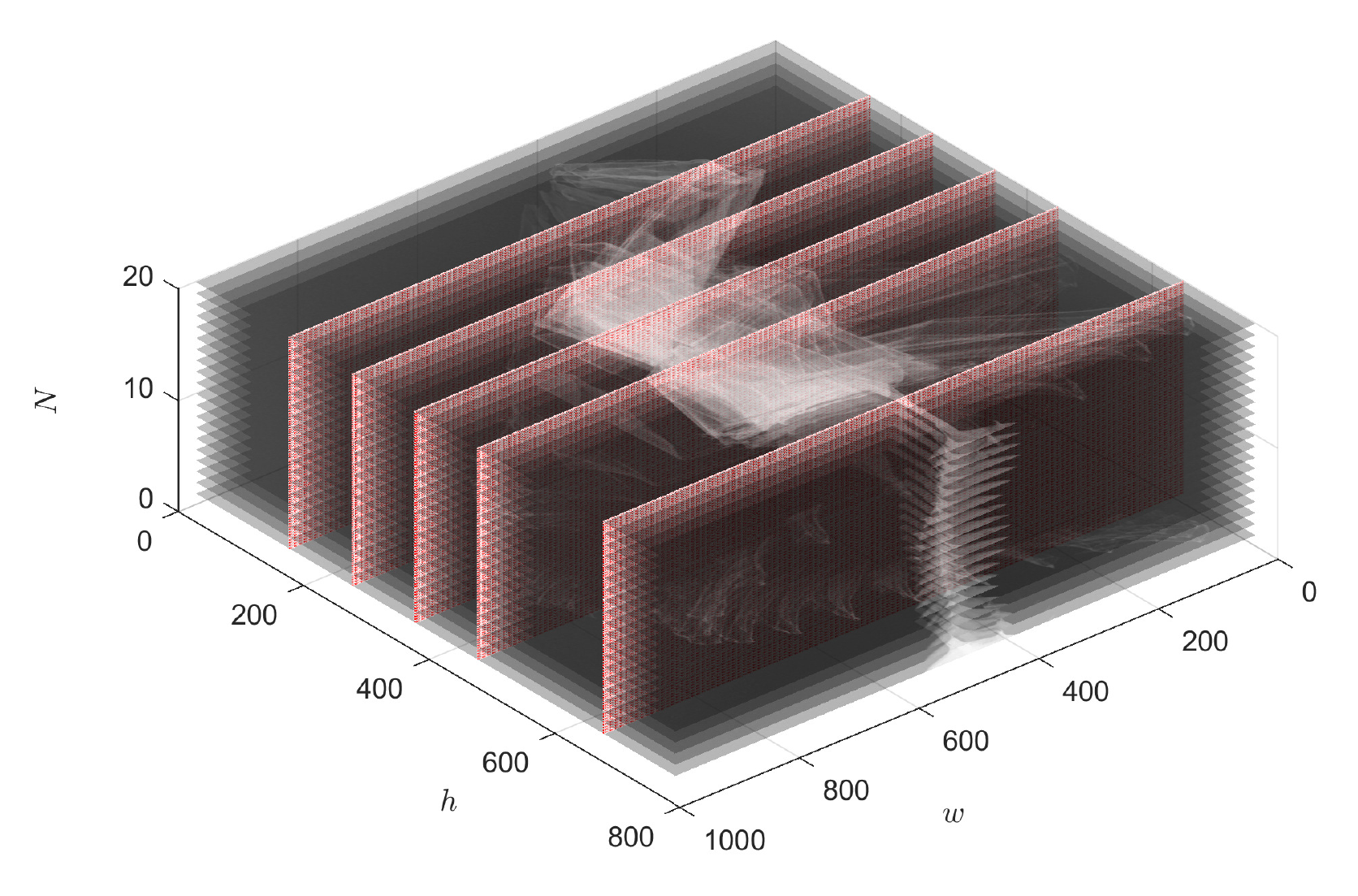}
\caption{Volumetric data of the crane. Visualized are 20 post-processed images and five cross-sections (red planes). }
\label{fig:postprob}
\end{figure}
\autoref{fig:postprob} shows part of the data and few of the horizontal cross-sections at the $h$ positions: $150, 250, 350, 450$ and $650,$ represented by the red planes. These sinograms are presented in \autoref{fig:sinograms} (from left to right).
\begin{figure}
\centering
\includegraphics[width=.8\linewidth]{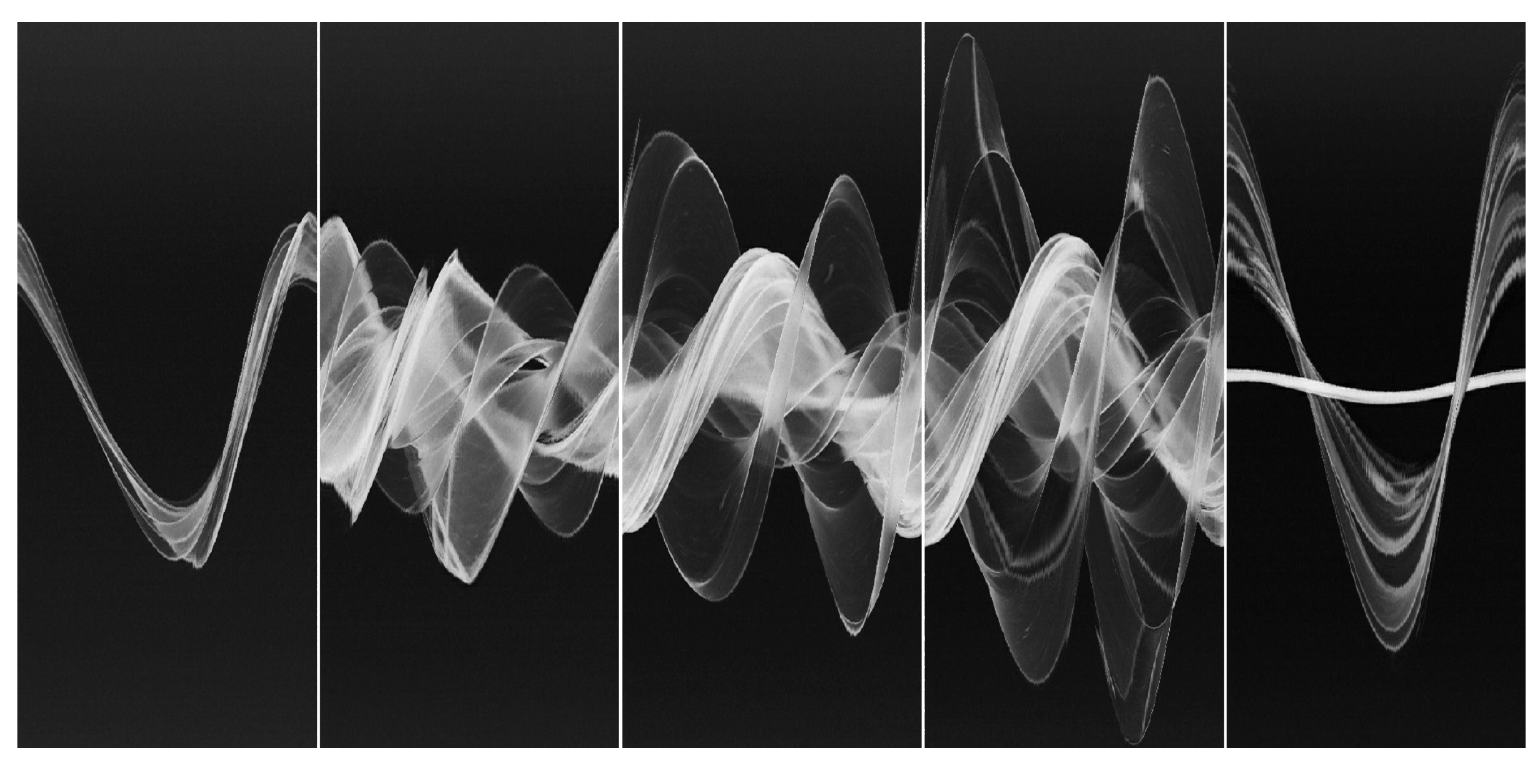}
\caption{The 2D sinograms are the horizontal cross-sections of the volumetric data. The images from left to right correspond to the red planes from top to bottom (as $h$ increases) presented in \autoref{fig:postprob}.}
\label{fig:sinograms}
\end{figure}

\begin{figure}
\centering
  \begin{subfigure}{.45\textwidth}
        \centering
        \includegraphics[width=.8\linewidth]{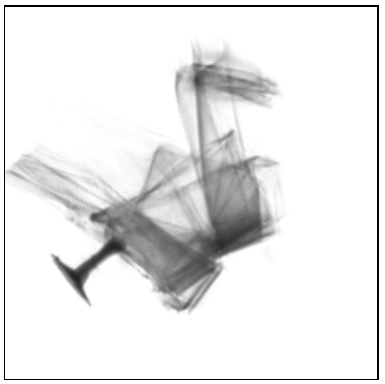}
    \end{subfigure}%
    ~ 
    \begin{subfigure}{.45\textwidth}
        \centering
        \includegraphics[width=.9\linewidth]{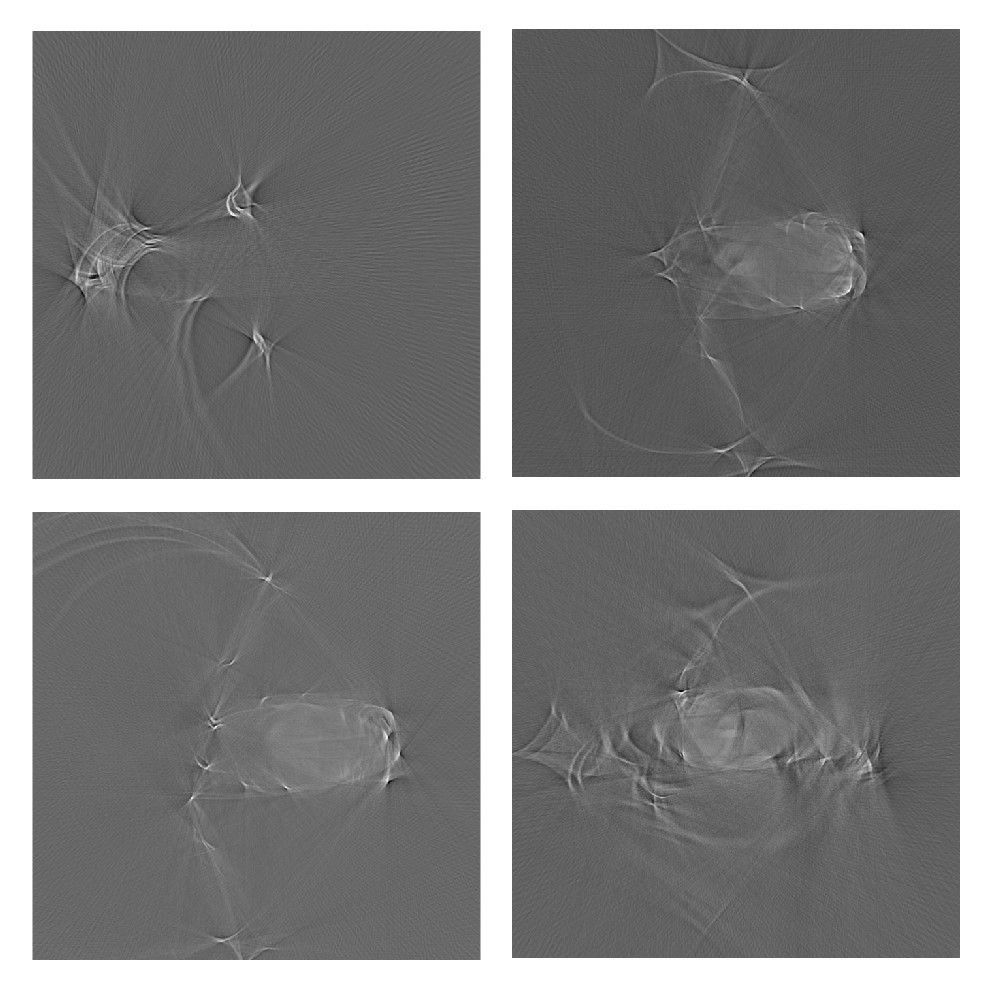}
    \end{subfigure}
    \caption{The reconstructed 3D origami crane from the simulated data (400 images). Print-screen of the volume raycaster (left) and its horizontal cross-sections at specific positions (right).}
\label{fig:raycaster}
\end{figure}

As in CTA scanning, we reconstruct the origami slice-by-slice. One of the most commonly used reconstruction methods is the filtered back-projection (see \cite{Kuc13} for more details). This algorithm is implemented in \texttt{Matlab} and used in the function \texttt{iradon}. We apply this routine to the 765 cross-section images and then the reconstructed object is the array of the 2D back-projected images. The recovered 3D origami can be visualized using a Raycaster. Its view at fixed angle and few horizontal cross-sections are presented in \autoref{fig:raycaster}. In \texttt{Matlab} language, it takes only a few lines of code to obtain the reconstructions. For convenience, we also included a fast evaluation of the filtered back-projection formula directly in ARTI.

\section{Modeling Cryo-microscopy}
``Cryo'' is a Greek word and means cold. It is used ambiguous in different areas of science: In Biology, Physics and Chemistry it refers to immobilizing a biological sample by ultracold freezing. 
As outlined in the introduction, this method can be used to visualize a particle (such as a molecule or a virus) in 3D from \emph{one} 2D micrograph. This is in fact theoretically possible if we assume that the micrograph visualizes several copies of the particle, which appear to be illuminated from different orientations, respectively. ``Cryo'' in Mathematics therefore often refers to a 3D reconstruction of an particle from 2D absorption images with unspecified illumination directions. 
At first glance this problem seems similar to CT, but the main difference, and the complication, is due to the fact that the orientation of every localized particle is unknown. 
Moreover, from a practical point of view, the level of noise in a micrograph is much higher compared to standard CT recordings. The complete problem of reconstruction a 3D particle is much too complicated for this short note. However, a micrograph can be relatively easily simulated
computationally based on data recorded with the remote setup shown in \autoref{fig:wobbling}. For a more rigorous mathematical approach, we refer to the recent papers \cite{BanCheLedSin20, KatKatSin15}.

\begin{figure}
\centering
    \includegraphics[scale=0.4]{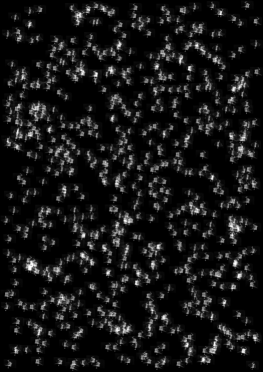}
    ~
    \includegraphics[scale=0.4]{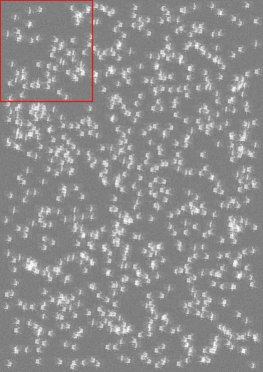}
    ~
    \includegraphics[scale=0.4]{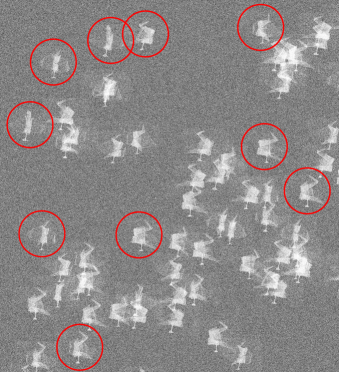}
    \caption{The sample image (simulated micrograph) (left) and its noisy version (middle). The right picture with the localized origamis corresponds to the zoomed-in version of the rectangle marked with red in the middle picture. }
\label{fig:micrograph}
\end{figure}

Computational methods for reconstructing 3D particles from micrographs are typically decomposed in three main steps as outlined below. There exist several open source packages, such as EMAN2 \cite{TanPenBal07} and ASPIRE \cite{Aspire}, where these steps are implemented.
\begin{enumerate}
\item The first step, called \emph{particle localization}, is to detect and extract localized 2D particles images from the micrograph. 
In the right picture of \autoref{fig:micrograph} we see the selected localized object images from our sample image (simulating a micrograph). 
\item Next comes the \emph{orientation estimation} step. Features of the 2D particle images are identified (for instance the beak of the origami crane). These features are used to determine the rotation angles. 
\item The last step is to visualize the particle in 3D from the 2D post-processed images. In practice, tens of thousands of these particles images are used as input for the reconstruction algorithm. \autoref{fig:circles} shows some of the extracted origami images from the sample image. 

\end{enumerate}

\subsection{The advanced origami scanner}

To simulate a micrograph we therefore need to create a sample image, which is a collection of 2D object images for arbitrary orientations of the origami. Such data cannot be created with the ``CTA'' origami scanner, presented in \autoref{sec_origami_scanner}, because it cannot rotate the particle off the central axis ($z$-direction). 

\begin{figure}
\centering
\includegraphics[width=.8\linewidth]{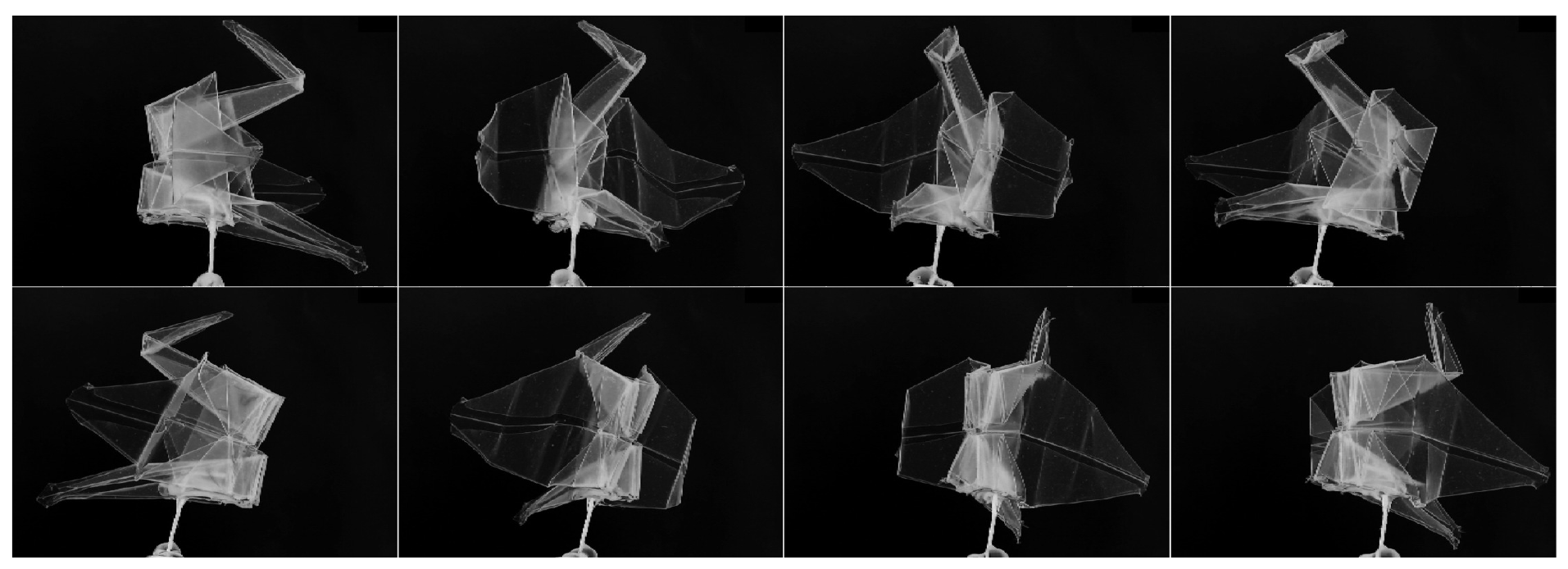}
\caption{Eight of the 800 post-processed images of the 3D origami crane at different orientations.}
\label{fig:pp_crane}
\end{figure}

However, the data recording and illumination devices remain the same as in the CTA setup. The only difference is the stepper motor, which is realized by placing the NEMA 17 stepper into a half tennis ball. The elasticity and the stickiness of its rubber inside fixates the stepper motor very well. This construction is placed on a plate above the base plate. Two additional stepper motors, called the $x$- and $y$-stepper, are fastened perpendicular to each other on the base plate and two pulleys are fastened on their axes. We tilt the half tennis ball with the $z$-stepper in the $x$-axis, using a wrapped string around the pulley of the $x$-stepper, attached to the opposite end of the half tennis ball. To fasten the string, under the elevated plate, inbound cuts are made and knots in the string. This assembly is shown in \autoref{fig:wobbling}.

\subsection{Cryo-EM data simulation and reconstructions}
Using this advanced setup, we generate $800$ 2D images of the origami crane for different orientations and we store them in a database. In \autoref{fig:pp_crane} we see eight of them.  The sample image consists of all object images placed at random positions (some overlapping might occur). Finally, we add noise with a signal-to-noise ratio (SNR) of $0.75$ and we create the noisy data, as shown in the middle picture of \autoref{fig:micrograph}.

\begin{figure}
\def\cutScale{0.25}
\centering
\includegraphics[scale=\cutScale]{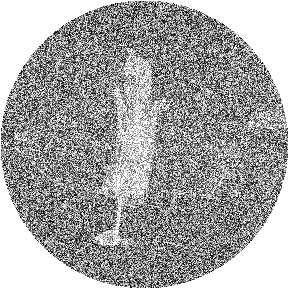}%
\includegraphics[scale=\cutScale]{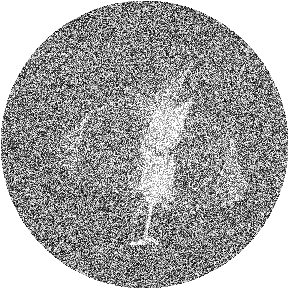}%
\includegraphics[scale=\cutScale]{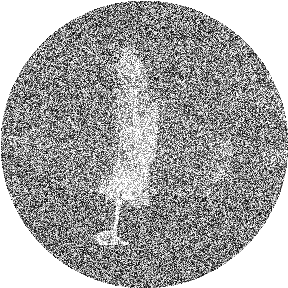}%
\includegraphics[scale=\cutScale]{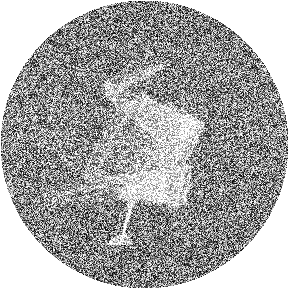}%
\includegraphics[scale=\cutScale]{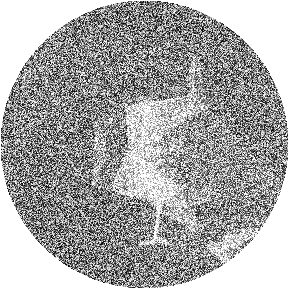}%
\\
\includegraphics[scale=\cutScale]{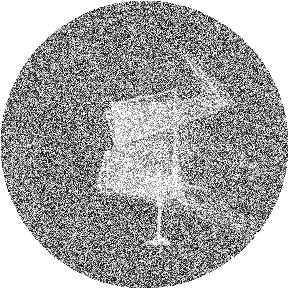}%
\includegraphics[scale=\cutScale]{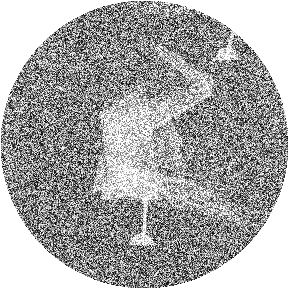}%
\includegraphics[scale=\cutScale]{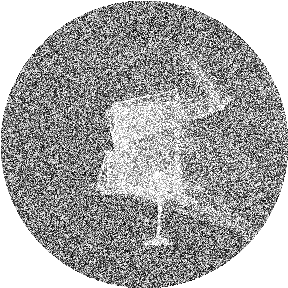}%
\includegraphics[scale=\cutScale]{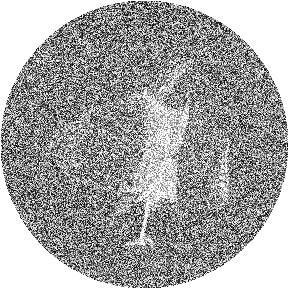}%
\includegraphics[scale=\cutScale]{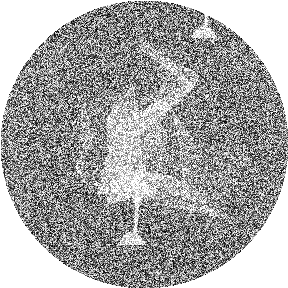}%
\caption{ The extracted object images from the marked area of the noisy sample image of \autoref{fig:micrograph}.}
\label{fig:circles}
\end{figure}

To give an idea of how complicated Cryo-imaging is, we apply a naive approach to reconstruct the 3D origami crane from the collected object images (see \autoref{fig:circles}). Note that in real applications, the noise level is much higher and the particles might not be identical (this makes the analysis even more difficult). This adds another statistical uncertainty and allows only for recovering a ``mean particle'' (averaging is performed). 

Theoretically a micrograph contains images of particles for every possible 3D rotation. We have seen already that data from rotation around one axis are sufficient for reconstructing a 3D object. Therefore, and this is what sometimes actually happens, one selects images, which are recorded from rotation around one axis only.  However, the rotation angle should be uniformly distributed in the remaining rotation direction in order to apply CTA imaging.

\begin{figure}
\centering
\includegraphics[width=.7\linewidth]{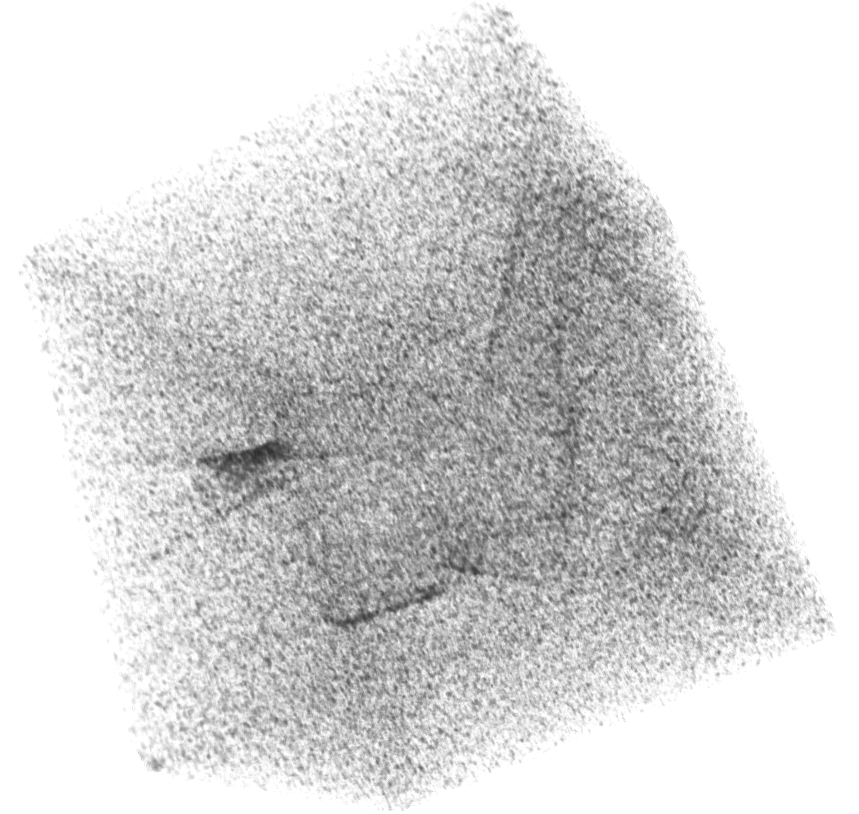}
\caption{The reconstructed 3D origami crane from the sample image (simulated Cryo-EM data).
}
\label{fig:cryoorigami}
\end{figure}

We use 80 origami images (extracted manually from the sample image consisting of 800 images) with known imaging directions (recovered from the filenames). In our case, the images correspond to rotation around one axis and the corresponding angles are almost uniformly distributed, such that we can still apply the \texttt{Matlab} function \texttt{iradon} to the collected data. A print screen of the volume raycaster of the reconstructed crane is presented in \autoref{fig:cryoorigami}. The quality is not satisfactory when compared with \autoref{fig:raycaster} but shows that even in this simplified and basic setting a reconstruction is feasible.

\appendix
\subsection*{Acknowledgements}
We thank Todd Quinto, Tufts University, for sharing his collection of literature on Allan MacLeod Cormack. OS acknowledges support from the Austrian Science Fund (FWF) within the SFB F68, project F6807-N36 (Tomography with Uncertainties) and I3661-N27 (Novel Error Measures and Source Conditions of Regularization Methods for Inverse Problems). Additionally, LM was supported by FWF, within the SFB F68, project F6801-N36. 

\section*{References}
\renewcommand{\i}{\ii}

\printbibliography[heading=none]

\end{document}